\documentclass[aps,prb,twocolumn,superscriptaddress,showpacs,floatfix,longbibliography]{revtex4-1}
\usepackage{amsmath,amssymb}

\usepackage[utf8]{inputenc}
\usepackage[T1]{fontenc}
\usepackage{xcolor}
\usepackage{booktabs}
\usepackage{array}
\IfFileExists{newtxtext.sty}
   {\usepackage{newtxtext,newtxmath}}
   {\IfFileExists{stix.sty}
      {\usepackage{stix}}
      {\IfFileExists{mathptmx.sty}
      {\usepackage{mathptmx}}{} } }

\usepackage{textcomp}

\usepackage{bm}

\IfFileExists{siunitx.sty}{\usepackage{booktabs,siunitx}}{}

\pdfoutput=1
\usepackage{color}
\definecolor{LinkColor}{rgb}{0.256,0.439,0.588}
\usepackage{hyperref}
\hypersetup{
   pdfauthor={abc},
   pdftitle={paper},
   colorlinks=true,
   citecolor=blue,
   linkcolor=blue,
   urlcolor=blue
}

\renewcommand{\vec}[1]{\mathbf{#1}}

\usepackage{pifont}

\usepackage[pdftex]{graphicx}
\DeclareGraphicsExtensions{.pdf,.jpeg,.png,.eps}
\usepackage{epstopdf}
\usepackage{soul}

\begin{document}
	\bibliographystyle{apsrev4-1}
	\title{Shubnikov-de Haas effect in the Falicov-Kimball model: strong correlation meets quantum oscillation}
	
	\author{Wei-Wei Yang}
	\affiliation{Key Laboratory of Quantum Theory and Applications of MoE $\&$ School of Physical Science and Technology, Lanzhou University, Lanzhou 730000, People Republic of China} %
	\author{Hong-Gang Luo}
	\affiliation{Key Laboratory of Quantum Theory and Applications of MoE $\&$ School of Physical Science and Technology, Lanzhou University, Lanzhou 730000, People Republic of China} %
	\affiliation{Beijing Computational Science Research Center, Beijing 100084, China}%
    \affiliation{Lanzhou Center for Theoretical Physics, Key Laboratory of Theoretical Physics of Gansu Province}

	\author{Yin Zhong}
    \email{zhongy@lzu.edu.cn}
    \affiliation{Key Laboratory of Quantum Theory and Applications of MoE $\&$ School of Physical Science and Technology, Lanzhou University, Lanzhou 730000, People Republic of China} %
    \affiliation{Lanzhou Center for Theoretical Physics, Key Laboratory of Theoretical Physics of Gansu Province}
	
\begin{abstract}
We present a comprehensive investigation of quantum oscillations (QOs) in the strongly-correlated Falicov-Kimball model (FKM). The FKM is a particularly suitable platform for probing the non-Fermi liquid state devoid of quasiparticles, affording exact Monte Carlo simulation across all parameter spaces.
In the high-correlation regime, we report the presence of prominent QOs in magnetoresistance and electron density at low temperatures within the phase separation state. The frequency behavior of these oscillations uncovers a transition in the Fermi surface as electron density diminishes, switching from hole-like to electron-like. Both types of Fermi surfaces are found to conform to the Onsager relation, establishing a connection between QOs frequency and Fermi surface area.
Upon exploring the temperature dependence of QOs amplitude, we discern a strong alignment with the Lifshitz-Kosevich (LK) theory, provided the effective mass is suitably renormalized. Notwithstanding, the substantial enhancement of the overall effective mass results in a notable suppression of the QOs amplitude within the examined temperature scope, a finding inconsistent with Fermi liquid predictions. For the most part, the effective mass diminishes as the temperature increases, but an unusual increase is observed at the proximity of the second-order phase transition instigated by thermal effects.
As the transition ensues, the regular QOs disappear, replaced by irregular ones in the non-Fermi liquid state under a high magnetic field.
We also uncover significant QOs in the insulating charge density wave state under weak interactions ($0 < U < 1$), a phenomenon we elucidate through analytical calculations.
Our findings shed light on the critical role of quasiparticles in the manifestation of QOs, enabling further understanding of their function in this context.
\end{abstract}
	
	\date{\today}
	
	\maketitle
	

\section{\label{sec1:level1}Introduction}	

When a magnetic field is applied to a system, Landau levels periodically intersect the Fermi level, resulting in oscillations of observables with respect to the inverse of the magnetic field \cite{QOs}. These quantum oscillations (QOs) are primarily caused by the oscillation of the density of states of quasiparticles. The most well-known oscillations are the Shubnikov–de Haas (SdH) effect and the de Haas-van Alphen (dHvA) effect, which correspond to the oscillation of magnetoresistance ($\rho_{xx}=\frac{\Delta R}{R_0}$) and thermodynamic magnetization, respectively. According to the Onsager relation \cite{QOs}, the frequency of QOs corresponds to the area of the extremal cross-section of the Fermi surface in momentum space perpendicular to the external magnetic field, which has been the standard techniques to extract Fermi surface in metallic quantum materials, e.g. the large-Fermi structure in overdoped cuprate high-$T_{c}$ superconductors \cite{Vignolle2008}. Furthermore, armed with classic Lifshitz-Kosevich (LK) theory \cite{QOs}, QOs have been successfully used to detect enhanced effective mass near quantum critical points in CeRhIn$_5$ and Ba(FeAs$_x$P$_{1-x}$)$_2$ \cite{doi:10.1143/JPSJ.74.1103,PhysRevLett.110.257002}. Therefore, QOs, combined with angle-resolved photoemission spectroscopy (ARPES), constitute the most useful method for measuring Fermi surface properties and uncover delocalization quantum phase
transition without symmetry breaking in CeCoIn$_{5}$ \cite{doi:10.1126/science.aaz4566}. Conventionally, the existence of QOs suggests a closed, well-defined, and coherent Fermi surface.

When considering electron correlation, Luttinger demonstrated that QOs can still be described by an LK-like theory as long as the Fermi liquid does not break down\cite{Luttinger1961}. This extended LK theory has been established within the framework of modern many-body perturbation theory \cite{Wasserman1996}.
The presence of QOs has generally been regarded as a clear indication of a Fermi-liquid state. However, the situation becomes more intricate in the presence of strong correlation effects. 
 QOs experiments in strongly correlated materials, such as high-$T_c$ cuprates \cite{PhysRevLett.100.047003,PhysRevLett.100.047004,PhysRevB.76.140508,PhysRevLett.100.187005,Doiron-Leyraud2007,doi:10.1146/annurev-conmatphys-030212-184305}, half-filled fractional quantum Hall system \cite{PhysRevLett.72.1906}, the exciton insulator \cite{Wang2021}, and the topological Kondo insulator SmB$_6$ and YbB$_{12}$ \cite{doi:10.1126/science.aaa7974,doi:10.1126/science.aap9607}, have led to perplexing outcomes.
In the underdoped regime of cuprates, QOs signals have been observed despite the well-established breakdown of the quasiparticle picture, as evidenced by the violation of the Luttinger theorem.
In the case of the topological Kondo insulator SmB$_6$, QOs imply the existence of a bulk neutral Fermi surface despite the insulating nature of bulk SmB$_6$, stimulating novel concepts such as the Majorana Fermi surface \cite{baskaran2015majorana,PhysRevB.102.155145}, surface Kondo breakdown \cite{PhysRevLett.116.046403},
and composite exciton Fermi liquid \cite{Chowdhury2018}.
(Note however that more conventional magnetic breakthrough-like mechanism seems to work \cite{PhysRevLett.115.146401,PhysRevLett.116.046404,PhysRevB.96.195122,PhysRevB.94.125140,doi:10.1073/pnas.2208373119,doi:10.1139/cjp-2022-0340}.)

The novel experimental findings and theoretical proposals mentioned above necessitate a reevaluation of the applicability of the LK-like theory in the presence of strong electron interactions.
However, even the investigation of basic models for electronic correlations, such as the Hubbard model in the presence of a magnetic field, requires sophisticated numerical techniques and QOs in the most interesting low-temperature regime remain unresolved.
Previous studies have employed various methods to tackle this challenge, including exact diagonalization \cite{PhysRevLett.77.4752,PhysRevB.74.125116}, as well as approximate approaches like Hartree-Fock mean-field theory \cite{PhysRevB.52.16744,PhysRevB.57.1312}, cluster perturbation theory \cite{Sherman2022}, renormalized mean-field theory in the large $U$ limit \cite{PhysRevB.97.035154}, dynamical mean-field theory (DMFT) \cite{PhysRevLett.127.196601,PhysRevB.100.115102,PhysRevB.96.235135}.
However, since QOs are typically observed at low temperatures, approximate methods like DMFT have limitations and cannot provide a comprehensive study of QOs at all temperatures.
Recently, a determinant quantum Monte Carlo (DQMC) algorithm has been applied to the Hubbard-Hofstadter model \cite{DQMC}. 
Similar to exact diagonalization, DQMC provides unbiased results, but a sign problem arises when deviating from half-filling, and thus QOs in the metallic quantum phase escape from the workhorse.

\begin{figure}
	\centering
	\includegraphics[width=0.9\columnwidth]{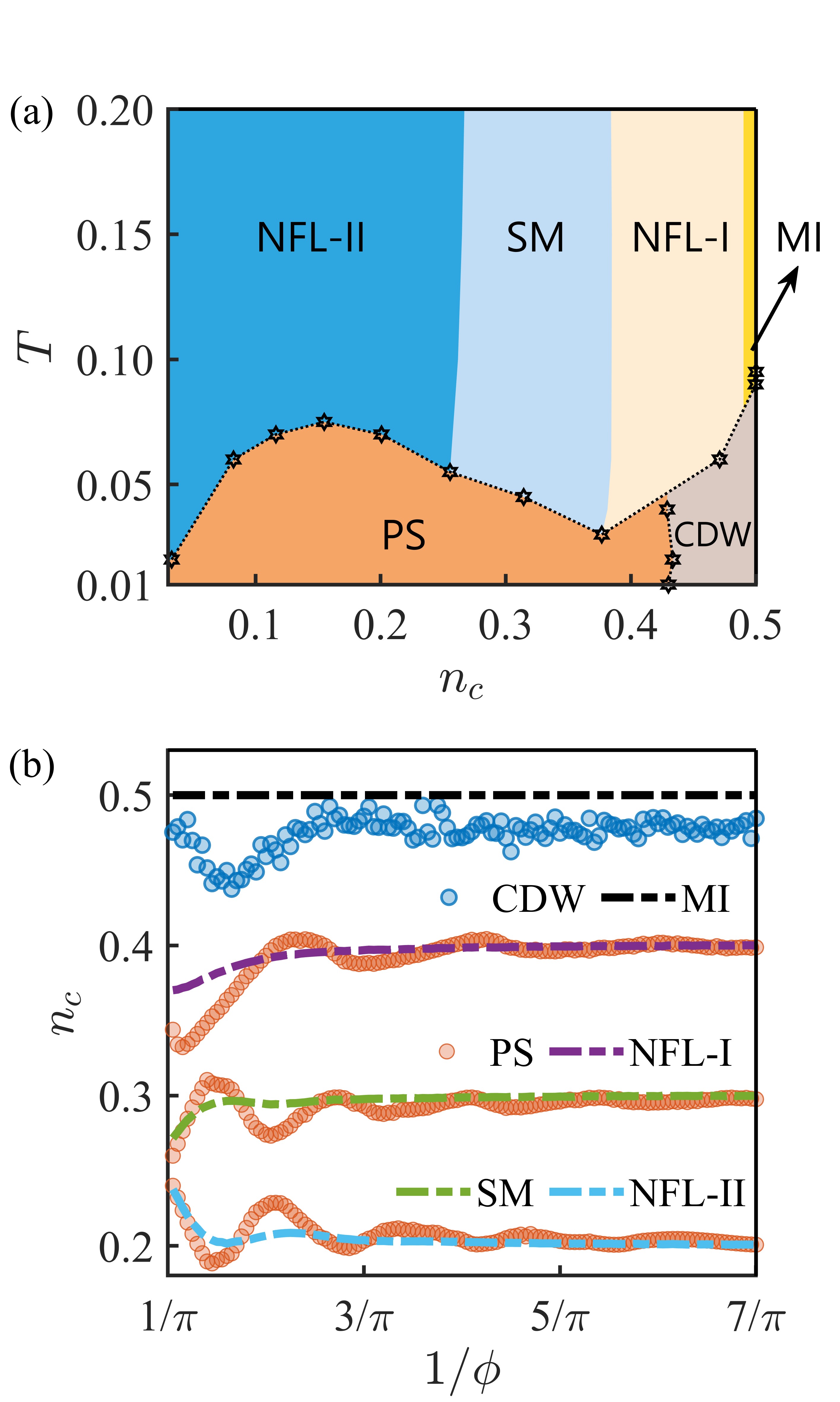}
	\caption{(a) The phase diagram of the Falicov-Kimball model (FKM) in the $n_c-T$ plane at $U=10$. The nearest neighbor hopping integral is used as the unit ($t=1$) to measure all energy scales. The low-temperature regime is divided into a charge density wave (CDW) state and phase separation (PS) state by a first-order transition. At high temperatures, there exists a Mott insulator (MI) state around the half-filling situation and three distinguishable non-Fermi liquid (NFL) states. With decreasing electron density, the MI crossover successively to the first non-Fermi liquid (NFL-I) state, the strange metal (SM) state, and finally to the second non-Fermi liquid (NFL-II) state. (b) The quantum oscillations (QOs) of electron density in different states of the doped FK model. The QOs occur only in the PS phase with various electron density.}
	\label{fig:global}
\end{figure}

To comprehensively analyze the temperature, doping, and interaction dependencies of QOs in strongly correlated systems, it is crucial to employ a suitable model.
 In this paper, we address this issue by utilizing the numerically solvable Falicov-Kimball model (FKM) to study QOs in strongly correlated scenarios\cite{RevModPhys.75.1333}.
The FKM offers several advantages for QOs research, as outlined below. Firstly, in our previous work \cite{PhysRevB.106.195117}, we extensively studied the doped FKM in the absence of a magnetic field using lattice Monte Carlo simulations. 
The FKM exhibits a rich variety of non-trivial phases away from half-filling, making it an ideal platform to explore the influence of correlations on QOs.  
Figure.~\ref{fig:global} (a) illustrates six distinct states observed, including a Mott insulator (MI), an insulator-like non-Fermi liquid (NFL-I), a strange metal (SM), a metallic non-Fermi liquid (NFL-II) at high temperatures, and a charge density wave (CDW) phase and phase separation (PS) at low temperatures. Notably, a violation of Luttinger's theorem is observed at high temperatures, indicating the absence of quasiparticles.
Moreover, employing unbiased Monte Carlo simulations enables us to investigate the FKM at any temperature $T$, coupling strength $U$, and magnetic field $B$. By simulating large-scale systems, we obtain high-resolution Fermi surface data, facilitating a detailed examination of the LK framework in the context of strongly correlated systems. Fig.~\ref{fig:global} (b) illustrates the overall electron density oscillations in the doped FKM.
Remarkably, QOs are prominently observed at low temperatures in the PS state. However, in the CDW state, although the particle number varies with the magnetic field, regular oscillations are not formed.
Interestingly, the FKM has also been employed to understand QOs in topological Kondo insulators using an exciton-like mean-field theory \cite{zyuzin2023haasvan}.
Overall, by utilizing the FKM, we bridge the gap in studying QOs under strong correlation effects and gain insights into their temperature, doping, and interaction dependencies. Through our analysis, we aim to shed light on the behavior of QOs in strongly correlated materials.

The remaining sections of this paper are structured as follows:
In Section \ref{sec2:level1}, we provide an introduction to the doped FKM in the presence of a magnetic field.
In Section \ref{sec3:level1}, we delve into the relationship between the frequency of QOs and the area of the Fermi surface as the electron density is varied.
In Section \ref{sec4:level1}, we focus on the temperature dependence of the QO amplitudes in both hole-like and electron-like Fermi surface scenarios. We employ the extended LK formula to extract the effective mass as a function of temperature.
In Section \ref{sec6:level1}, we investigate QOs in the low-temperature insulating CDW state under weak interactions and provide an analytical computation to explain the origin of these exotic QOs.
Finally, in Section \ref{sec5:level1}, we provide a summary of our findings and conclusions.
By organizing the paper in this manner, we aim to present a comprehensive analysis of QOs in the doped FKM, accounting for the effects of magnetic fields, electron density variations, and temperature dependencies.

\section{\label{sec2:level1}The model}
We consider the FKM \cite{RevModPhys.75.1333} on a two-dimensional square lattice (in the $xy$ plane) with a lattice constant $a=1$, the Hamiltonian without external magnetic field is defined as
\begin{equation}
	\hat{H}(w)=-t\sum_{m}(\hat{c}_{m}^{\dag}\hat{c}_{m\pm\vec{x}}
	+\hat{c}_{m}^{\dag}\hat{c}_{m\pm\vec{y}})
	+U\sum_{m}\hat{n}_{m}\hat{w}_{m}-\mu\sum_{m}\hat{n}_{m}.
	\label{eq:eff_model0}
\end{equation}
Here $\hat{c}_{m}^{\dag}(\hat{c}_{m})$ is the itinerant electron's creation (annihilation) operator at site $m$.
The index $m\pm\vec{x}(\vec{y})$ denotes the neighbor site of $m$ in $x(y)$ direction. In this context, $t$-term denotes the hopping integral and only nearest-neighbor-hoping is involved.
$\hat{n}_{m}=\hat{c}_{m}^{\dag}\hat{c}_{m}$ is the particle number of the itinerant electron, while $\hat{w}_m$ is the particle number operator for the electron in the localized state. $U$ is the onsite Coulomb interaction between the itinerant electron and the localized electron. The hole doping into the half-filled system ($\mu=\frac{U}{2}$) is realized by tuning the chemical potential $\mu$.

By Peierl's substitution \cite{Peierls1,PhysRevB.81.115119} the effect of a uniform magnetic field $\vec{B}=B\vec{e}_z$ in the $z$ direction can be induced by introducing a site-dependent phase factor in the hopping integral $t_{ij}=t \times e^{i\frac{2\pi}{\phi_0} \int_{R_i}^{R_j}A(\vec{r}) \cdot dr} $, where the unit flux quanta $\phi_0=h/e$.
$\vec{A}$ is the vector potential, and the Hamiltonian is defined as
\begin{equation}
	\begin{aligned}		
	\hat{H}=&-t\sum_{m}(e^{ia_{m,m\pm\vec{x}}}\hat{c}_{m}^{\dag}\hat{c}_{m\pm\vec{x}}
	+e^{ia_{m,m\pm\vec{y}}}\hat{c}_{m}^{\dag}\hat{c}_{m\pm\vec{y}})\\
	&+U\sum_{m}\hat{n}_{m}\hat{w}_{m}-\mu\sum_{m}\hat{n}_{m},
		\label{eq:model1}
	\end{aligned}
\end{equation}
In Landau gauge we assume $\vec{A}=Bx\vec{e}_y$, the phase can be simplified as $a_{m,m\pm\vec{x}}=0$ and $a_{m,m\pm\vec{y}}=\pm\frac{e}{\hbar}Bm_xa^2$. Here, $m_x$ is the $x$ coordinate of site $m$, and $\phi=Ba^2$ is the magnetic flux in a plaquette.
In this paper, the study of this square lattice FKM with open boundary conditions is carried out with Monte Carlo simulation on a system as large as $N_s=20\times20$. 
To measure all energy scales, we adopt the nearest neighbor hopping integral as the unit ($t=1$), and for the sake of numerical simulation convenience, we set $e=\hbar=a=1$. As a result, our model can be simplified as follows:
\begin{equation}
	\begin{aligned}		
	\hat{H}=&-\sum_{m}(\hat{c}_{m}^{\dag}\hat{c}_{m\pm\vec{x}}
	+e^{\pm i\phi m_x}\hat{c}_{m}^{\dag}\hat{c}_{m\pm\vec{y}})\\
	&+U\sum_{m}\hat{n}_{m}\hat{w}_{m}-\mu\sum_{m}\hat{n}_{m}.\nonumber
	\end{aligned}
\end{equation}

To attack the QOs involving the strong correlation effect, we focus on situation $U=10$.

\begin{figure}
	\centering
	\includegraphics[width=0.8\columnwidth]{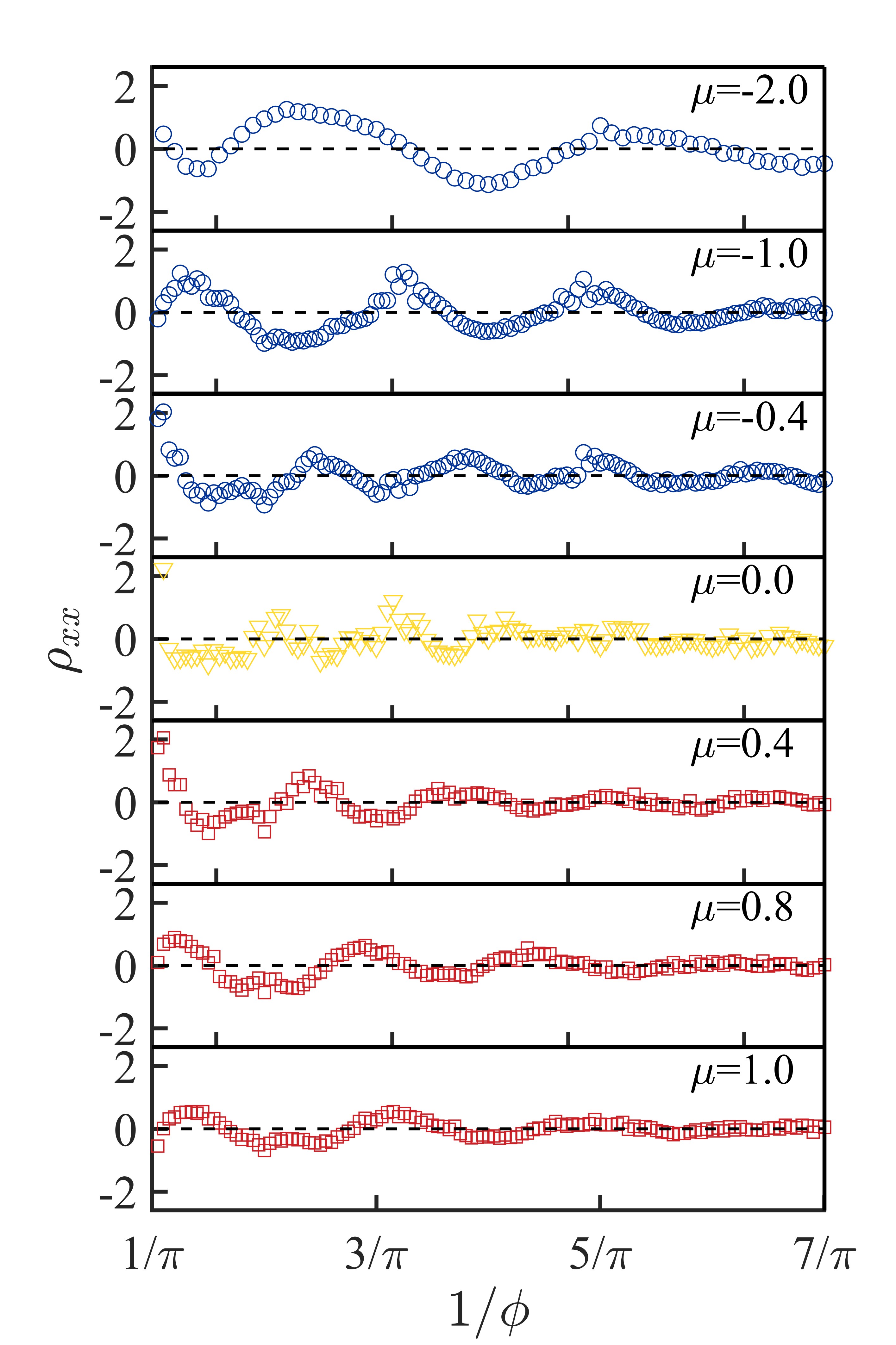}
	\caption{The SdH QOs with varying magnetic field in the PS state. The red square denotes a hole-like Fermi surface whereas the blue circle denotes a electron-like one. The yellow triangle denotes the data around the nesting of Fermi surface, where the QOs is not obvious. The non-oscillating background is subtracted using exponential fits to obtain the QOs component of the magnetoresistance. Error bars are smaller than the size of data points.}
	\label{fig:freq1}
\end{figure}


\section{\label{sec3:level1}frequency}

\begin{figure}
	\centering
	\includegraphics[width=1\columnwidth]{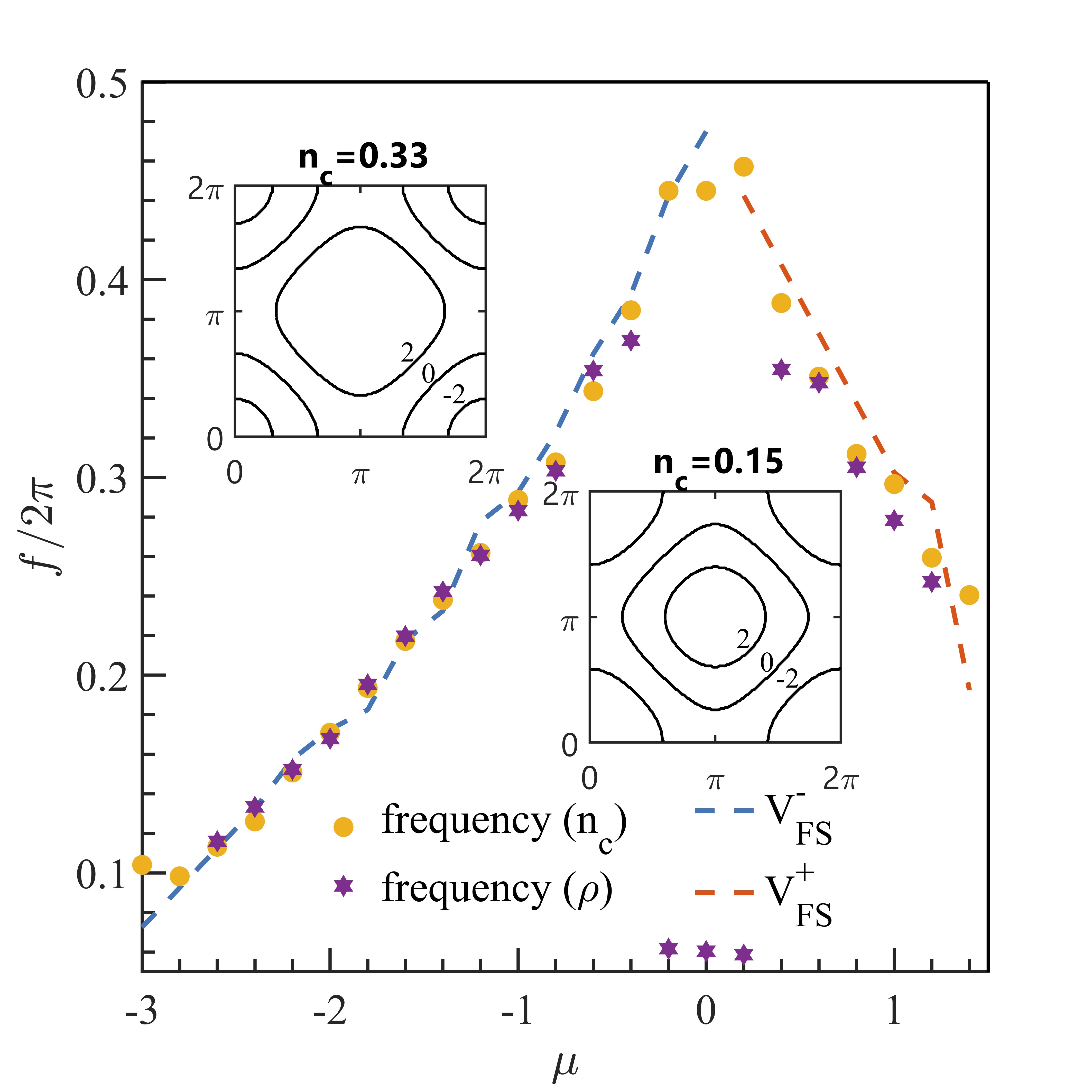}
	\caption{The $\mu-$dependent frequency of the Shubnikov–de Haas (SdH) QOs (purple star) and the QOs of electron density $n_c$ (yellow circle). Away from the nesting of Fermi surface ($\mu \sim 0$), the frequency is closely agreeing with the area of Fermi surface. Inset shows the constant energy surface for both hole-like Fermi surface ($n_c=0.33$) and electron-like Fermi surface ($n_c=0.15$). With decreasing $n_c$, the hole-like Fermi surface (blue dashed line) changes to the electron-like Fermi surface (red dashed line). With large $n_c$ ($\mu \gtrsim 1.4$), the system transits to the CDW state. The QOs disappear in the insulating CDW state. Plots of the QOs are measured at $T=0.03$. For convenience, both the area of Fermi surface is scaled by the area of the whole Brillouin zone ($V_\text{FS}/(2\pi)^2$).}
	\label{fig:freq}
\end{figure}

\begin{figure}
	\centering
	\includegraphics[width=0.8\columnwidth]{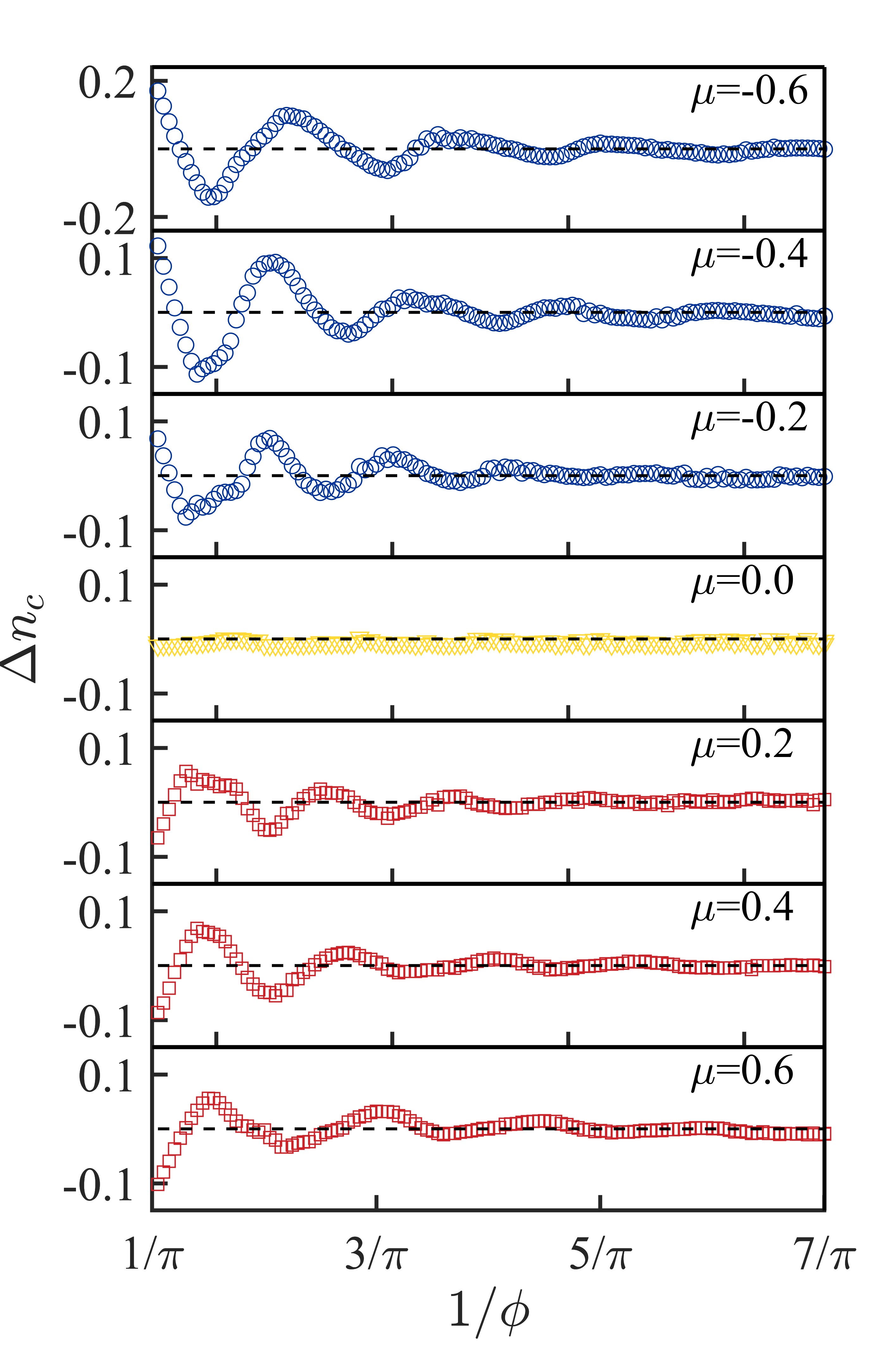}
	\caption{The electron density QOs with varying magnetic field in the PS state. The red circle denotes a hole-like Fermi surface whereas the blue square denotes a electron-like one. The yellow triangle denotes the data around the nesting of Fermi surface, where the QOs are not obvious. The non-oscillating background is subtracted using exponential fits to obtain the oscillating component of the electron density $\Delta n_c$. Error bars are smaller than the size of data points.}
	\label{fig:f3}
\end{figure}

In conventional metals, QOs are typically observed in the low-temperature regime, which is inversely proportional to the effective mass. In the FKM, the presence of strong interaction ($U/t=10$) leads to a significant enhancement of the effective mass, restricting the occurrence of QOs to an even lower temperature regime. In the low$-T$ regime of the doped FKM, two distinct states are present: the CDW and the PS state \cite{PhysRevB.106.195117}. 
The previous investigations on cuprates have highlighted the importance of having a unified Fermi surface, such as the Fermi pocket, for the occurrence of QOs.
These QOs exhibit a dominant frequency and carriers that are associated with the closed and coherent Fermi surface
 \cite{PhysRevLett.100.187005,PhysRevLett.108.196403,doi:10.1126/science.aay7311,Doiron-Leyraud,PhysRevLett.100.047003,https://doi.org/10.1038/nphys2792}. This observation is also corroborated within the FKM, where QOs exclusively manifest in the metallic states that sustain an intact Fermi surface. Figure \ref{fig:freq1} illustrates the prominent SdH oscillations at $T=0.03$ in the PS regime, corresponding to the cooling-down phase of the SM and NFL-I (red square), as well as the cooling-down phase of NFL-II (blue circle), across a wide range of magnetic fields. Consistent with prior research \cite{QOs,Wasserman1996}, the insulator-like CDW state exhibits no regular oscillatory behavior (see Fig.~\ref{fig:global} (b)).

Under a strong magnetic field, the Fermi surface undergoes deformation. Our study establishes a correspondence between the frequency of SdH oscillations and the Fermi surface in the absence of a magnetic field. In Fig.~\ref{fig:freq}, we extract the SdH oscillation frequencies (purple stars) at $T=0.03$ using fast Fourier transform within a magnetic field range of $\phi \in [\frac{\pi}{7},\pi]$.  Similarly, the electron density $n_c$ exhibits clear oscillatory behavior with a corresponding frequency, as shown by the yellow circles in Fig.~\ref{fig:freq}.
The oscillation of electron density, defined as $\Delta n_c=\frac{n_c(\vec{B})-n_c(\vec{0})}{n_c(\vec{0})}$, where $n_c(\vec{0})$ represents the electron density in the absence of a magnetic field, follows a pattern similar to the magnetoresistance. The dashed line represents the Fermi surface area without a magnetic field, which has been scaled by the area of the Brillouin zone ($V_\text{FS}/(2\pi)^2$) for convenience. The frequency is directly related to the Fermi surface area $V_\text{FS}$ through the Onsager relation $f=\frac{\hbar}{2\pi} V_\text{FS}$.
Importantly, in this hole-doped MI, the frequency does not exhibit a monotonic decrease with decreasing electron density. Instead, there is a turning point around $\mu=0, n_c=0.25$. The inset in the figure provides a visualization of the constant energy surface of the spectrum for $n_c=0.33$ (hole-like Fermi surface) and $n_c=0.15$ (electron-like Fermi surface). This indicates that during the cooling-down regime of the SM and NFL-I states ($\mu>0, n_c>0.25$), holes are likely to act as the predominant mobile carriers. Conversely, with an increase in hole concentration ($\mu<0, n_c<0.25$), electrons become the dominant mobile carriers. These findings suggest that the Onsager relation holds true irrespective of whether the Fermi surface is hole-like or electron-like.

There are two additional observations to note in Fig.~\ref{fig:freq} and Fig.~\ref{fig:freq1}.
Firstly, the disappearance of QOs coincides with the transition of the Fermi surface from the hole-like one to the electron-like one. This phenomenon arises due to the interplay between hole-like and electron-like excitations, which can be elucidated by examining the oscillatory behavior of the electron density, as depicted in Fig.~\ref{fig:f3}. On different sides of the Fermi surface nesting, the magnetic field induces opposite variations in the electron density.  When the Fermi surface approaches the nesting condition, both particle-like and hole-like excitations contribute with equal magnitudes but opposite signs, leading to the suppression of oscillations. In the system with predominately hole-like excitations ($\mu>0, n_c>0.25$), the observed opposite oscillation tendencies between resistivity and electron density are well-founded. As the excited particles, the holes, which carry the current, increase with decreasing electron density.
Furthermore, although the frequencies of both the electron density ($n_c$) oscillations and the SdH oscillations are proportional to $V_\text{FS}$, there exists a deviation between them. This deviation can be rationalized by considering that the SdH oscillations primarily originate from states near the Fermi surface, while all eigen-energies contribute to the $n_c$ oscillations. Therefore, it is expected that the two frequencies would not be exactly the same, resulting in the observed discrepancy.

\begin{figure}
	\centering
	\includegraphics[width=0.9\columnwidth]{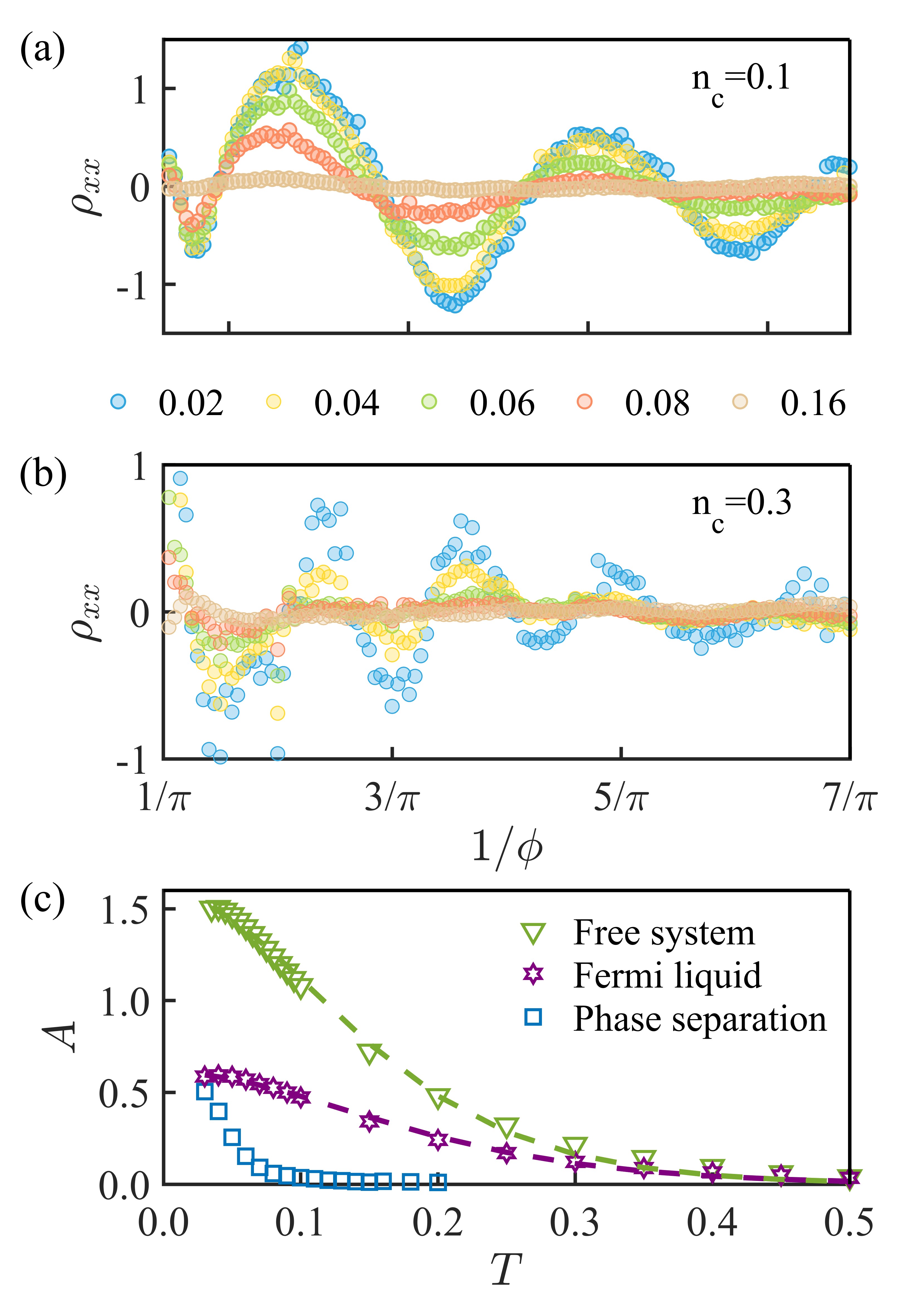}
	\caption{(a) QOs of magnetoresistance measured at various temperatures with $n_c(\vec{0})=0.1$. These curves are measured at $T=0.02, 0.04, 0.06, 0.08, 0.16$, respectively. (b) QOs of magnetoresistance measured at various temperatures with $n_c(\vec{0})=0.3$ measured at the same temperatures. To extract the oscillatory component, a smooth and nonoscillatory background is removed from the data. (c) Temperature dependence of the oscillation amplitude $A(T)$ for the free system, the Fermi-liquid and the PS state ($n_c(\vec{0})=0.3$). The dashed line are fits to the LK formula.}
	\label{fig:AT}
\end{figure}

\section{\label{sec4:level1}amplitude}

\begin{figure}
	\centering
	\includegraphics[width=0.9\columnwidth]{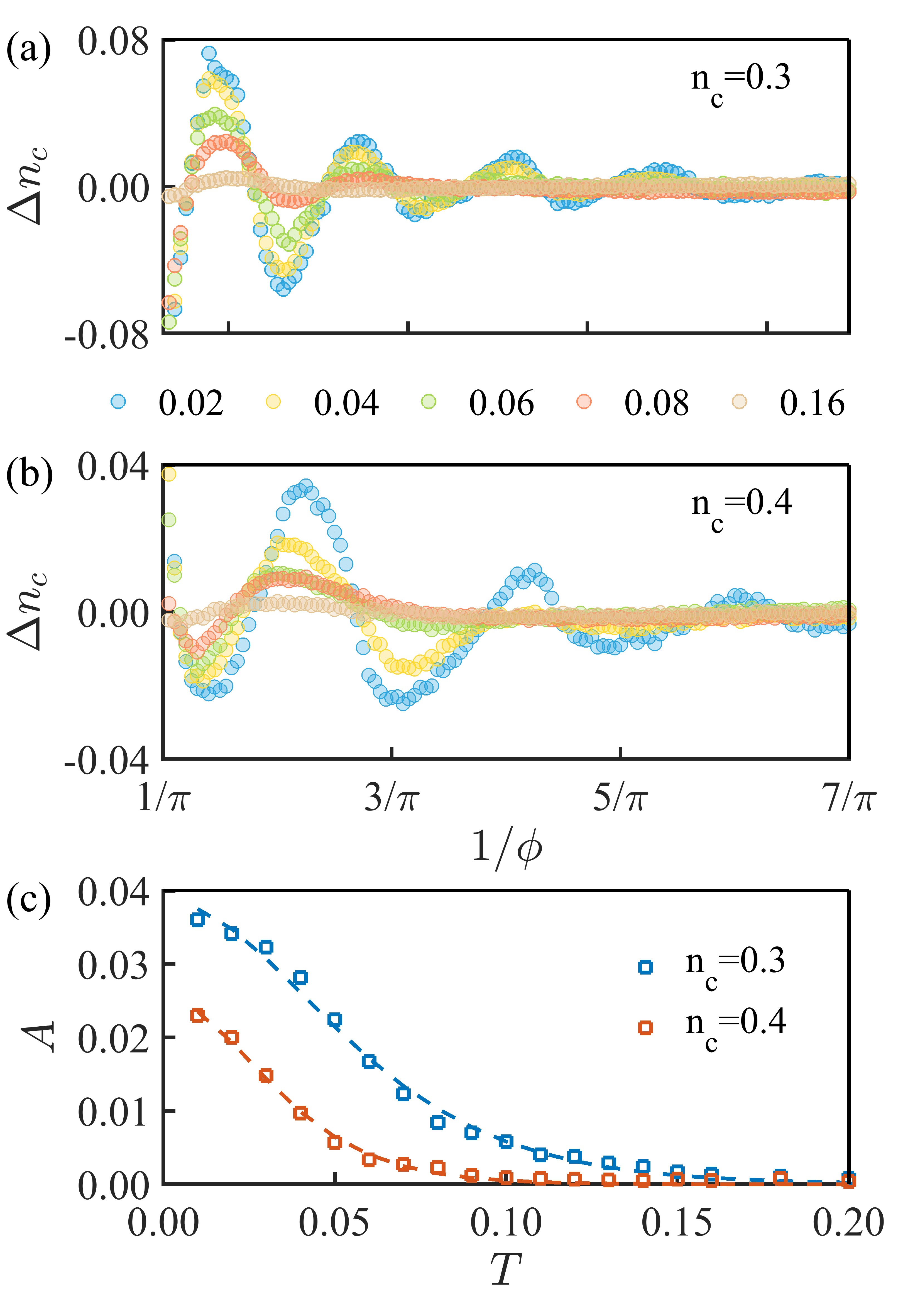}
	\caption{Electron density QOs in the system with hole-like excitations. $\Delta_{n_c}$ with varying magnetic field with (a) $n_c(\vec{0})=0.3$ and (b) $n_c(\vec{0})=0.4$.  (c) QOs amplitude as a function of temperature. The dashed line is the best fit to the LK formula.}
	\label{fig:AT3}
\end{figure}

\begin{figure}
	\centering
	\includegraphics[width=0.9\columnwidth]{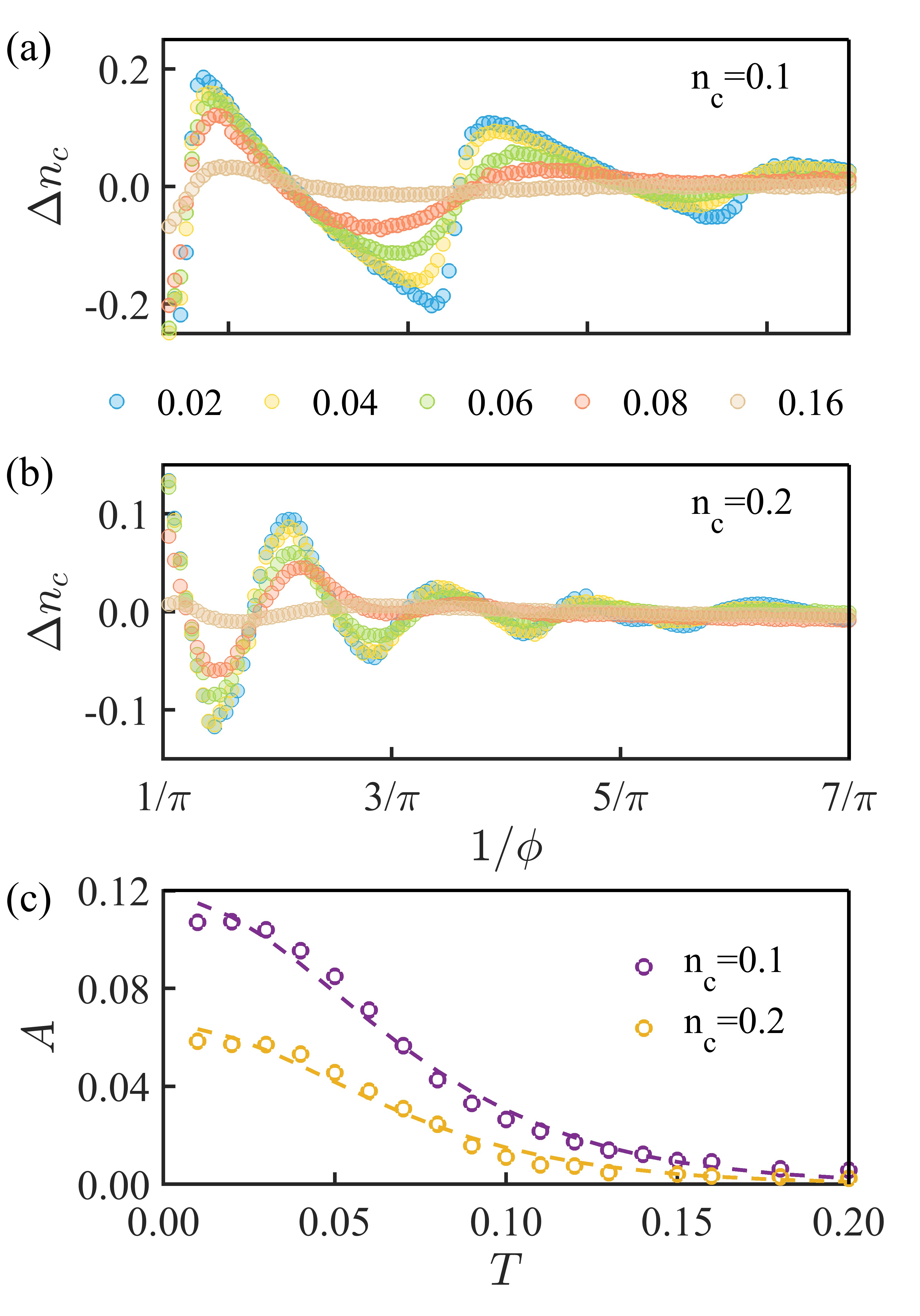}
	\caption{Electron density QOs in the system with electron-like excitations. $\Delta_{n_c}$ with varying magnetic field with (a) $n_c(\vec{0})=0.1$ and (b) $n_c(\vec{0})=0.2$.  (c) QOs amplitude as a function of temperature. The dashed line is the best fit to the LK formula.}
	\label{fig:AT4}
\end{figure}

 To investigate the temperature dependence of the QOs, we examine the amplitude $A(T)$ while keeping the particle number fixed. As mentioned earlier, the frequency is directly related to the Fermi surface area $V_\text{FS}$ through the Onsager relation. Since $V_\text{FS}$ and the electron density $n_c$ are uniquely related to each other at different temperatures, it is necessary to fix $n_c(\vec{0})$ when studying the temperature dependence of the amplitude $A(T)$.
 In Fig.~\ref{fig:AT}, we present the temperature dependence of the SdH oscillations with a fixed particle number. Figure.~\ref{fig:AT} (a) corresponds to the cooling-down regime of the NFL-II state ($n_c=0.1$), while Figure.~\ref{fig:AT} (b) represents the cooling-down regime of the SM state ($n_c=0.3$). The magnetoresistance is plotted over a range of magnetic fields and various temperatures, with a subtracted non-oscillatory background.
 Figure.~\ref{fig:AT} (c) displays the temperature dependence of $A(T)$ for the oscillations observed in Fig.~\ref{fig:AT} (b), extracted from the fast Fourier transform. For SdH oscillations, a monotonic decay as the temperature increases is observed, which is expected in conventional metals. However, the vanishing slope at zero temperature is not evident, as the amplitude is heavily suppressed at low temperatures ($T<0.03$, not shown here) due to the upturn of resistance $R_0$ caused by weak localization in the two-dimensional system.
 For comparison, we include the $A(T)$ curves for conventional metals, including a free system (green triangle, $U=0$) and a Fermi liquid (purple star, $U=1$). Both conventional metals have the same itinerant electron concentration ($n_c=0.3$).
 Remarkably, although the overall $T$-dependent amplitude of Fermi liquid is suppressed compared with the free system, both of them exhibit similar trends and almost sustain up to the same high temperature. These trends are well described by the LK formula:
 \begin{equation}
 	A_0(T)=\frac{X}{\sinh X},
 	\label{eq:AT}
 \end{equation}
 where $X=2\pi^2(k_BT)/(\hbar \omega_c)$. Here, $\omega_c=eB/m^{\ast}$ represents the cyclotron frequency, $k_B$ is the Boltzmann constant, $\hbar$ is the reduced Planck constant, and $m^{*}$ denotes the effective mass of the carriers \cite{Lifshitz-Kosevich}. The dashed line corresponds to the best fit of Eq.~(\ref{eq:AT}). The similar temperature dependence observed for the free system and the Fermi liquid suggests that their effective masses are comparable.
In contrast, the enhanced effective mass in the $U=10$ case leads to a significant suppression of both the overall amplitude and the temperature range supporting QOs. Specifically, for the $n_c=0.3$ scenario, the regular SdH oscillations can only be sustained up to $T/t \approx 0.06$, indicating exclusive presence of regular QOs in the PS state.

 \begin{figure}
 	\centering
 	\includegraphics[width=0.6\columnwidth]{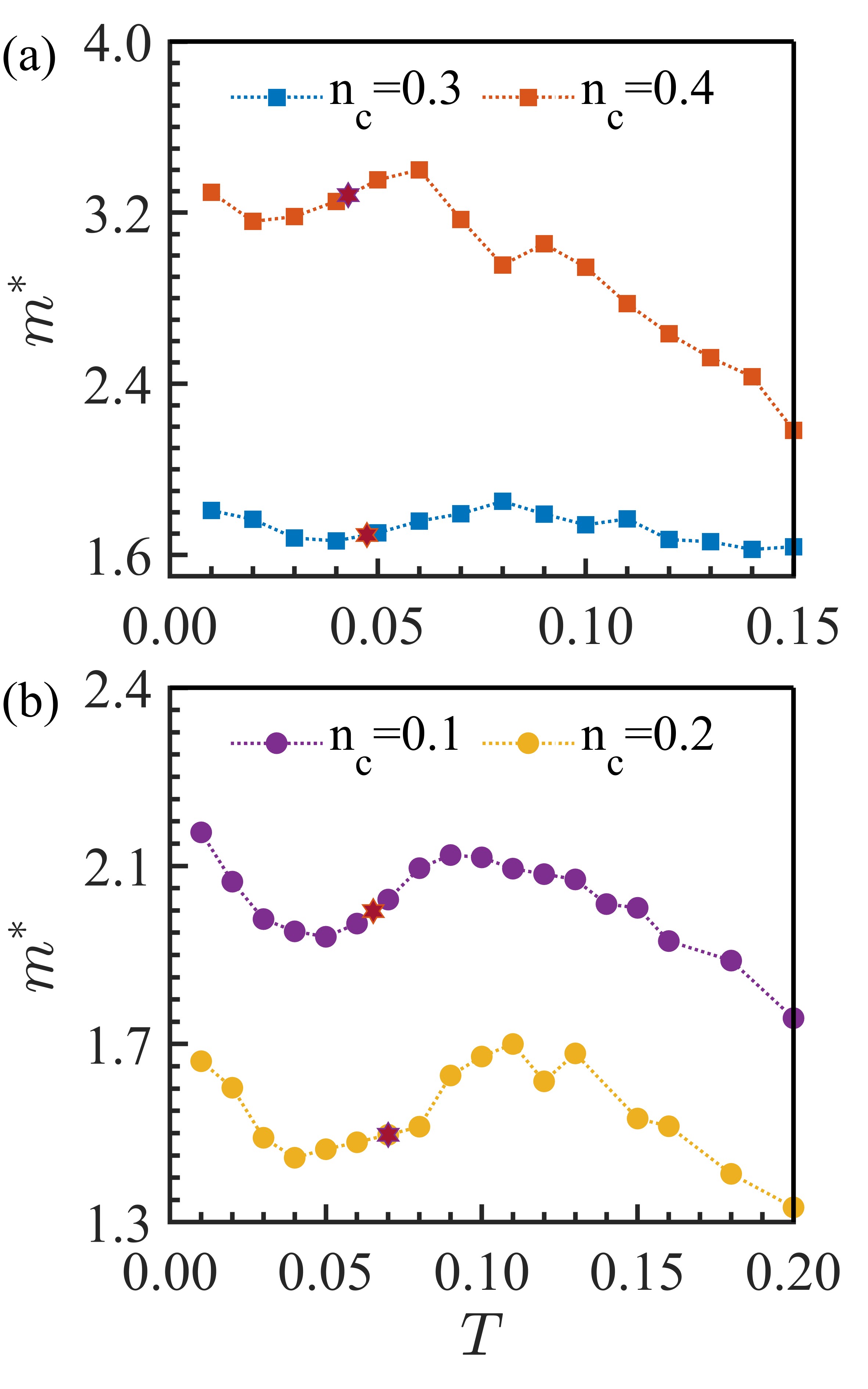}
 	\caption{Effective mass $m^{*}$ versus temperature for the system with (a) hole-like excitation and (b) electron-like excitation.
 	The temperature of thermal-driven second order phase transition is marked with the red star.}
 	\label{fig:Me}
 \end{figure}

In our previous work, we demonstrated that the Hubbard-I approximation is effective in the FKM \cite{PhysRevB.106.195117}. Within the Hubbard-I analysis, at high temperatures, the self-energy $\Sigma$ can be described by the expression $	\Sigma(\omega_m)=\frac{U^2/4}{i\omega_m+\mu-\frac{U}{2}}$, and $\text{Im}\Sigma(\omega_m)=-\omega_m\frac{U^2/4}{\omega^2_m+(\mu-\frac{U}{2})^2}$.
 According to the extended LK theory\cite{Wasserman1996}, the amplitude in a correlated system can be expressed as follows:
 \begin{equation}
 	A_p(T)=\frac{(2\pi)^2pT}{\hbar \omega_c} \sum_{\omega_m>0} e^{-\frac{2\pi p}{\hbar \omega_c} (\omega_m-\text{Im}\Sigma(\omega_m))}.
 	\label{eq:AT2}
 \end{equation}
Under the assumption that the imaginary part of the self-energy can be written as $\text{Im}\Sigma(\omega_m) = \omega_m \times f(\omega_m)$, the effect of correlation on the $A(T)$ can be understood as a renormalization of the effective mass, i.e.,
 \begin{equation}
	A_p(T)=\frac{(2\pi)^2pT}{\hbar \omega_c} \sum_{\omega_m>0} e^{-\frac{2\pi p}{\hbar \omega_c^{*}} (\omega_m)},
	\label{eq:AT2}
\end{equation}
where $\omega_c^{*}=\frac{eB}{m^{*}}$ and $m^{*}=m_e(1-f(\omega_m))$.
Here, for the Hubbard-I approximation $m^{*}=m_e\left(1+\frac{U^2/4}{\omega^2+(\mu-\frac{U}{2})^2}\right)$.
However, to reproduce the characteristic $\frac{X}{\text{sinh}X}$ shape of the amplitude curve, an analytical function $f(\omega_m)$ of $\omega_m$ is enough, which is not restricted to the specific form given by the Hubbard-I formula.
In Fig.~\ref{fig:AT3} (a,b), the electron density oscillations in the system with hole-like excitations are plotted. Figure.~\ref{fig:AT3} (c) shows the amplitudes extracted by applying the fast Fourier transform to the data as a function of temperature. The dashed lines represent the best-fitting data to the extended LK formula for $A(T)$.
Figure.~\ref{fig:AT4} illustrates the electron density oscillations in the system with electron-like excitations, and the corresponding amplitudes as a function of temperature. The dashed lines represent the best-fitting data to the extended LK formula for $A(T)$, showing excellent agreement with the original data.
The behavior of $A(T)$ for electron-like excitations is similar to that of hole-like excitations, except that the former exhibits a larger amplitude. This difference can be attributed to the smaller value of $n_c(\vec{0})$ in the electron-like case.

In strongly correlated systems, the effective mass often exhibits a temperature dependence, typically decreasing with increasing temperature. As mentioned before, the LK formula, which describes the behavior of the amplitude, effectively captures the characteristics of strongly correlated PS states by considering a renormalized effective mass.
To further analyze this, we extracted the effective mass $m^*$ as a function of temperature. Figure.~\ref{fig:Me} (a) and Figure.~\ref{fig:Me} (b) correspond to the cases of hole-like and electron-like excitations, respectively. The red star denotes the critical temperature $T_c$ of the thermally driven phase transition.
It is noteworthy that weak, irregular QOs persist as the system transitions into the low-temperature regime of NFL states. 
This observation allows us to observe the trend of effective mass variations in the slightly higher temperature region, which is the reason for the wide range of temperature axis in Fig.~\ref{fig:Me}.
By comparing Fig.~\ref{fig:Me} with the SdH oscillations demonstrated in Fig.~\ref{fig:AT}, we observe that the overall enhancement of $m^{*}$ accounts for the suppression of QOs at high temperatures. Interestingly, unlike conventional Fermi liquids, the temperature dependence of $m^{*}$ is not a monotonous function. Notably, a contrasting trend emerges around the PS-NFL phase transition (see stars in Fig.~\ref{fig:Me}). Deep within the NFL and PS states, the decreasing tendency of $m^{*}$ persists, suggesting that the phase transition can lead to an increase in the effective mass.


\section{\label{sec6:level1}quantum oscillation in the insulator state}

\begin{figure}
	\centering
	\includegraphics[width=0.8\columnwidth]{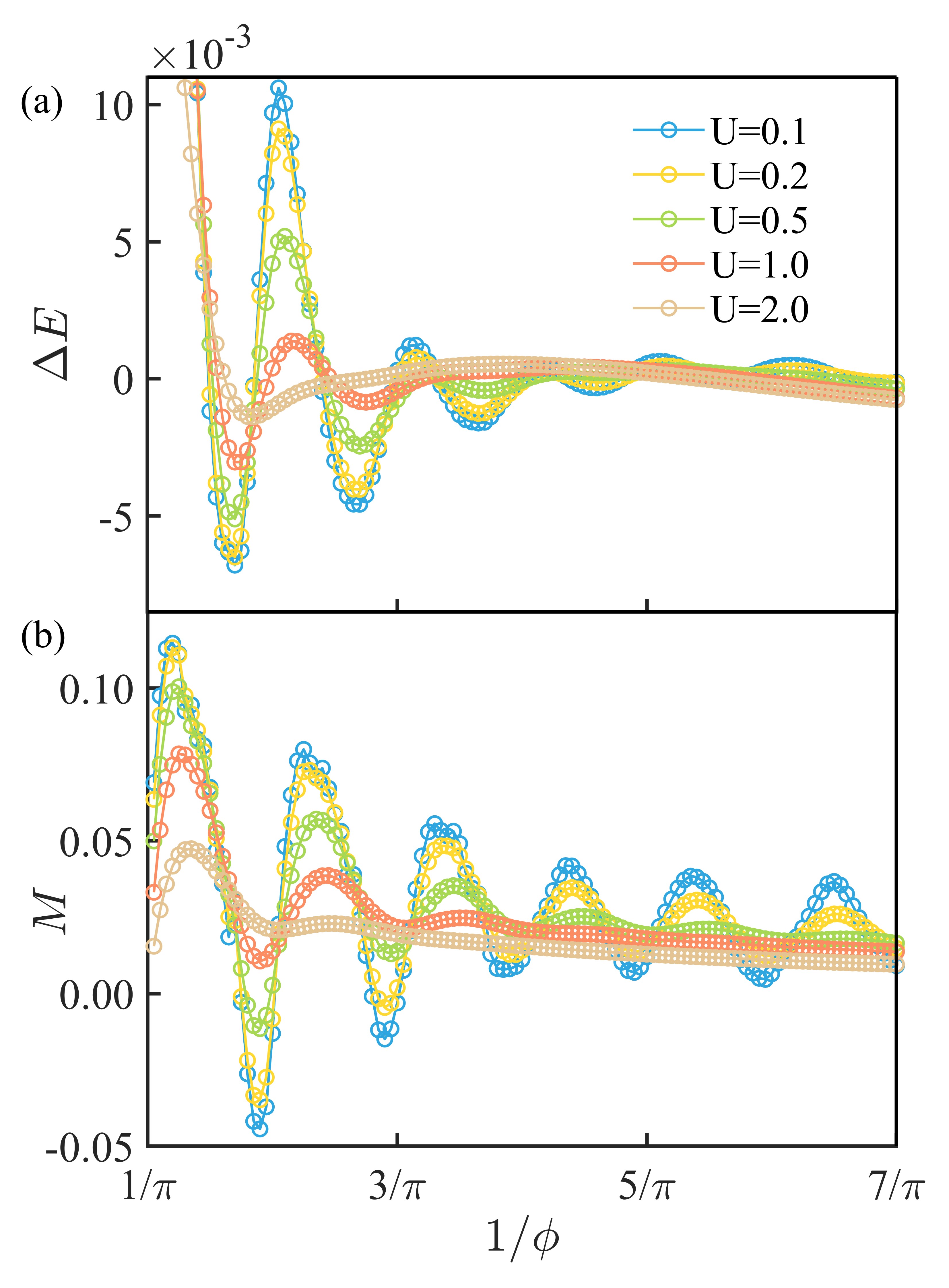}
	\caption{QOs of (a) energy $\Delta E$ and (b) magnetization ($M=-\frac{\partial E}{\partial B}$) with varying magnetic field at different correlation strength $U$.
	The non-oscillating background is subtracted using exponential fits to obtain the oscillating component of $\Delta E$.
	The magnetization exhibits no pronounced overall increase or decrease trend as the magnetic field varies, thus no additional processing or adjustments are applied.}
	\label{fig:CDW}
\end{figure}

Previous studies have revealed that the FKM exhibits a CDW state at low temperatures when the system is at half-filling ($\mu=\frac{U}{2}$). Interestingly, we have observed the presence of QOs in this insulating state when the correlation is weak ($0<U<1$).
Figure.~\ref{fig:CDW} demonstrates that for weak coupling situations, both the energy $E$ and magnetization $M=-\frac{\partial E}{\partial B}$ exhibit periodic variations with changes in the magnetic field. 
Moreover, as the interaction strength increases, the amplitude of the QOs decreases. When the interaction strength reaches $U=2$, QOs completely disappear. However, when the interaction $U < 0.5$, clear QOs are observed despite the insulating properties. The oscillation frequency is nearly the same as that observed in the weak coupling regime, approximately $f \sim \pi$, indicating a Fermi surface with a size of half the Brillouin zone.
This suggests that the observed frequency is consistent with the frequency in the absence of a band gap, indicating that the Fermi surface measured by QOs in the CDW state corresponds to the Fermi surface of the metallic state before the formation of the insulating state.

Below, we will analytically explain the exotic QOs observed in the insulating state.
At zero temperature and half-filling, the FKM is equivalent to a lattice model with a staggered potential $\Delta (-1)^{m_x+m_y}\hat{c}^{\dagger}_m\hat{c}_m$, as proven by Kennedy and Lieb \cite{KENNEDY1986320}. This potential introduces a gap of $2\Delta$ in the insulating state. 
In this case, $\Delta=\frac{U}{2}$ and the Hamiltonian can be written as:
\begin{equation}
	\hat{H}=-t\sum_{m}(\hat{c}_{m}^{\dag}\hat{c}_{m\pm\vec{x}}
	+\hat{c}_{m}^{\dag}\hat{c}_{m\pm\vec{y}})
	+\Delta \sum_m (-1) ^{m_x+m_y} \hat{c}^{\dagger}_m \hat{c}_m.
	\label{eq:eff_model_CDW1}
\end{equation}
By the Fourier transformation,
\begin{equation}
	\begin{aligned}	
	\hat{H}&=\sum_k \epsilon_k \hat{c}_k^{\dagger} \hat{c}_k +\Delta \sum_k \hat{c}_k^{\dagger} \hat{c}_{k+Q}\\
	&=\sum_k^{'} ( \epsilon_k \hat{c}_k^{\dagger} \hat{c}_k +\epsilon_{k+Q} \hat{c}_{k+Q}^{\dagger} \hat{c}_{k+Q} )+\Delta \sum_k^{'} (\hat{c}_k^{\dagger} \hat{c}_{k+Q}
	+\hat{c}_{k+Q}^{\dagger} \hat{c}_{k}),\\
	\label{eq:eff_model0_CDW2}
	\end{aligned}	
\end{equation}
where $\epsilon_k=-2t(\text{cos}k_x+\text{cos}k_y)$, $Q=(\pi,\pi)$. Here, $\sum_{k}^{'}$ denotes the summation of $k$ is limited in the reduced Brillouin zone,
i.e., the regime where $\text{cos}k_x+\text{cos}k_y>0$.
Since $\epsilon_k=-\epsilon_{k+Q}$, the dispersion relation is $E_{k\pm}=\pm \sqrt{\epsilon_k^2+\Delta^2}$.
At half-filling, the electrons completely fill the states up to the position where $\epsilon_k=0$ in the energy band.
Due to the formation of two intersecting energy bands by electrons with momenta of $k$ and $k+Q$, a hybridization between these two bands opens up a gap of size $2\Delta$ at the position of $\epsilon_k=0$. 
As a result, an insulating state with two energy bands, $E_{k\pm}$, is established, with $E_{k-}$ being fully occupied at half-filling.
The frequency of QOs that appear in the insulating CDW state corresponds to the Fermi surface existed at $\epsilon_k=0$ before the opening of the energy gap due to hybridization.

The oscillation of energy with magnetic field can be directly calculated by the following analytical expression.
Under the staggered potential and the orbital magnetic field, the Hamiltonian of the CDW state can be written as
\begin{equation}
	\begin{aligned}	
		\hat{H}=-t\sum_{m}(\hat{c}_{m}^{\dag}\hat{c}_{m+\vec{x}}
		+e^{ i\phi m_x} \hat{c}_{m}^{\dag}\hat{c}_{m+\vec{y}}+\text{H.c.})\\
		+\Delta \sum_m (-1) ^{m_x+m_y} \hat{c}^{\dagger}_m \hat{c}_m.
		\label{eq:eff_model0_CDW3}
	\end{aligned}	
\end{equation}
For the metallic state ($\Delta=0$), under the magnetic field the dispersion relation $\epsilon_k=-2t(\text{cos}k_x+\text{cos}k_y)$ becomes discrete Landau levels $\epsilon_n=-\epsilon_0+(n+\frac{1}{2})\hbar \omega_{c} $, where $n=0,1,2,...$ and $-\epsilon_0$ is the energy of the band bottom.
We define $B_+$ as the magnetic field that aligns the $(n-1)$th Landau level with the Fermi energy, and $B$ as the magnetic field that aligns the $n$th Landau level with the Fermi energy. Hence, the relationship between the magnetic field and the Landau levels can be expressed as:
\begin{equation}
	\begin{aligned}	
		-\epsilon_0+(n+\frac{1}{2})\hbar \frac{eB}{m}=0,\\
		-\epsilon_0+(n-1+\frac{1}{2})\hbar \frac{eB_+}{m}=0.\\
		\label{eq:eff_model0_CDW4}
	\end{aligned}	
\end{equation}
Thus, the periodicity is confirmed by
\begin{equation}
	\begin{aligned}	
 \frac{1}{B}-\frac{1}{B_+}=\frac{\hbar e }{m\epsilon_0},\\
		\label{eq:eff_model0_CDW5}
	\end{aligned}	
\end{equation}
where $\frac{\hbar e }{m\epsilon_0}=\frac{2 \pi e}{\hbar}\frac{1}{A_F}$ and $A_F=\pi k_F^2$ is the volume of the 2-dimensional Fermi surface.

On the other hand, in the insulating CDW state ($\Delta \neq 0$) under a magnetic field, the discrete Landau levels are given by $E_{n\pm}=\pm \sqrt{\epsilon_n^2+\Delta^2}$.
Obviously, there exists no Landau level at the Fermi energy.
However, at the bottom of the upper band ($E_{n+}=+\Delta$) and the top of the lower band ($E_{n-}=-\Delta$),
the Landau level could be occupied.
Regarding the top of the upper band, the relationship between the magnetic field and the Landau levels can be expressed as:
\begin{equation}
	\begin{aligned}	
		-\epsilon_0+(n+\frac{1}{2})\hbar \frac{eB}{m}=-\Delta,\\
		-\epsilon_0+(n-1+\frac{1}{2})\hbar \frac{eB_+}{m}=-\Delta.\\
		\label{eq:eff_model0_CDW6}
	\end{aligned}	
\end{equation}
Therefore, Equation~\ref{eq:eff_model0_CDW5} remains valid in the CDW state, which suggests a same frequency with the metallic state.

Next, we can analytically calculate the amplitude of the QOs in the CDW state.
At zero temperature and zero magnetic field, the energy is $E=\sum_{k}^{'}E_{k-}=-\sum_{k}^{'}\sqrt{\epsilon_k^2 +\Delta^2}$.
At the presence of magnetic field, the summation $\sum_{k}^{'}$ transforms to $D\sum^{\infty}_{n=0}$, ($D$ is number of states in each Landau level) thus
\begin{equation}
	\begin{aligned}	
		E=-D\sum^{\infty}_{n=0} \sqrt{\left(-\epsilon_0+(n+\frac{1}{2})\hbar \omega_{c}\right)^2 +\Delta^2}.
		\label{eq:eff_model0_CDW7}
	\end{aligned}	
\end{equation}
By using the summation formula $\sum^{\infty}_{n=0} f(n+\gamma)=\int_{0}^{\infty} dx f(x) +2 \int_{0}^{\infty}dx f(x) \text{Re}\sum^{\infty}_{p=1} e^{i2\pi p(x-\gamma)}$, we have
\begin{equation}
	\begin{aligned}	
		E=&-D\int_{0}^{\infty} dx \sqrt{\left(-\epsilon_0+x\hbar \omega_{c}\right)^2 +\Delta^2}\\
		&-2D \int_{0}^{\infty} dx \sqrt{\left(-\epsilon_0+x\hbar \omega_{c}\right)^2 +\Delta^2} \text{Re} \sum^{\infty}_{p=1} e^{i2\pi p(x-\frac{1}{2})}.
		\label{eq:eff_model0_CDW8}
	\end{aligned}	
\end{equation}
Only the second term will cause oscillations in the energy as a function of magnetic field. Thus, the oscillatory part can be written as
\begin{equation}
	\begin{aligned}	
		E_{\text{osc}}=
		-2D \hbar \omega_{c} \int_{0}^{\infty} dx \sqrt{\left(x-\frac{\epsilon_0}{\hbar \omega_{c}}\right)^2 +\left(\frac{\Delta}{\hbar \omega_{c}}\right)^2} \\
\times\text{Re} \sum^{\infty}_{p=1} (-1)^p e^{i2\pi px}.
		\label{eq:eff_model0_CDW9}
	\end{aligned}	
\end{equation}
By making a variable substitution $x^{'}=x-\frac{\epsilon_0}{\hbar \omega_c}, \alpha=\frac{\Delta}{\hbar \omega_{c}}$, we obtain
\begin{equation}
	\begin{aligned}	
	E_{\text{osc}}=
	-2D \hbar \omega_{c} \text{Re} \sum^{\infty}_{p=1} (-1)^p e^{i2\pi p\epsilon_0/\hbar \omega_{c}} \\
	 \times \int_{-\epsilon_0/\hbar \omega_{c}}^{\infty} dx^{'} \sqrt{\alpha^2+x^{'2}} e^{i2\pi p x^{'}}.
		\label{eq:eff_model0_CDW10}
	\end{aligned}	
\end{equation}
By partial integration
\begin{equation}
	\begin{aligned}	
		&\int_{-\epsilon_0/\hbar \omega_{c}}^{\infty} dx^{'} \sqrt{\alpha^2+x^{'2}} e^{i2\pi p x^{'}} =\\
		&\frac{\sqrt{\alpha^2+x^{'2}} e^{i2\pi p x^{'}}|^{\infty}_{-\epsilon_0/\hbar \omega_{c}}}{i2\pi p}-\frac{1}{i 2 \pi p} \int_{-\epsilon_0/\hbar \omega_{c}}^{\infty} dx^{'} \frac{x^{'}}{\sqrt{\alpha^2+x^{'2}}}e^{i 2\pi p x^{'}},
		\label{eq:eff_model0_CDW11}
	\end{aligned}	
\end{equation}
where the first term does not produce oscillations and can therefore be omitted.
The second term can be written in the form of a derivative with respect to $p$, thus
\begin{equation}
	\begin{aligned}	
		&\int_{-\epsilon_0/\hbar \omega_{c}}^{\infty} dx^{'} \sqrt{\alpha^2+x^{'2}} e^{i2\pi p x^{'}} = \\
		&-\frac{1}{ (i2 \pi)^2 p} \frac{d}{dp}\int_{-\epsilon_0/\hbar \omega_{c}}^{\infty} dx^{'} \frac{e^{i 2\pi p x^{'}}}{\sqrt{\alpha^2+x^{'2}}}+...
		\label{eq:eff_model0_CDW12}
	\end{aligned}	
\end{equation}
When the magnetic field is weak, i.e., $\epsilon_0=4t >>\hbar \omega_{c}$, the lower limit of integration can be extended to $-\infty$, and thus
$\int_{-\infty}^{\infty} dx^{'} \frac{e^{i 2\pi p x^{'}}}{\sqrt{\alpha^2+x^{'2}}}=K_0(\alpha 2 \pi p)$.
Here, $K_0(x)$ is the modified Bessel function of the second kind. Thus Eq.~\ref{eq:eff_model0_CDW10} is written as
\begin{equation}
	\begin{aligned}	
		E_{\text{osc}}&=
		-2D \hbar \omega_{c} \text{Re} \sum^{\infty}_{p=1} (-1)^p e^{i2\pi p\epsilon_0/\hbar \omega_{c}} \frac{1}{ (2 \pi)^2 p} \frac{d}{dp} K_0(\alpha 2 \pi p) \\
		&=\frac{D\Delta}{\pi}  \sum^{\infty}_{p=1} \frac{(-1)^p}{p} \text{cos} \left( \frac{2\pi p \epsilon_0}{\hbar \omega_{c}} \right) K_1 \left( \frac{2\pi p \Delta}{\hbar \omega_{c}} \right).
		\label{eq:eff_model0_CDW13}
	\end{aligned}	
\end{equation}
The observed QOs exhibit a period that aligns with the previous analysis and is determined by $\epsilon_0$.  In the absence of a band gap, $\epsilon_0$ represents the position of the Fermi energy relative to the bottom of the band. Therefore, it confirms that the falsely measured Fermi surface in the insulating CDW state corresponds to the Fermi surface of the metal at $\epsilon_k=0$.
In the limit where the gap $\Delta \gg \hbar \omega_{c}$, employing the asymptotic form of $K_1(x) \sim \frac{\pi}{2\sqrt{x}}e^{-x}$, we ultimately obtain
\begin{equation}
	\begin{aligned}	
		E_{\text{osc}} \simeq
		\frac{D\Delta}{\pi}  \sum^{\infty}_{p=1} \frac{(-1)^p}{p} \text{cos} \left( \frac{2\pi p \epsilon_0}{\hbar \omega_{c}} \right) \frac{\pi}{2 \sqrt{\frac{2\pi p \Delta}{\hbar \omega_{c}}}}
		e^{-\frac{2\pi p \Delta}{\hbar \omega_{c}}}.
		\label{eq:eff_model0_CDW14}
	\end{aligned}	
\end{equation}
It turns out that the behavior of the QOs amplitudes in the insulating CDW state aligns with the expectations of the LK formula.
The presence of a band gap acts as a Dingle factor $e^{-\frac{2\pi p \Delta}{\hbar \omega_{c}}}$, so QOs can only be observed when the band gap is sufficiently small.

All in all, the above numerical and analytical analysis reinforces the idea that insulating states can indeed have QOs \cite{PhysRevLett.115.146401,PhysRevLett.116.046404,PhysRevB.96.195122,PhysRevB.94.125140,doi:10.1073/pnas.2208373119,doi:10.1139/cjp-2022-0340},
which is stimulated by intensive experimental studies in topological Kondo insulator SmB$_{6}$ and YbB$_{12}$. The contribution of our work here is that the QOs in insulating CDW states have been confirmed by exact Monte Carlo simulation, which go beyond previous studies involving only non-interacting quasiparticles.

Before ending this section, it is noted that the FKM has been used to understand QOs in topological Kondo insulator with a exciton-like mean-field theory \cite{zyuzin2023haasvan}. However, we should emphasize that the insulating state of FKM (without dispersion of localized electron) at half-filling corresponds to a CDW state rather than the exciton-like insulator predicted by mean-field theory. Therefore, the QOs inferred from exciton-like mean-field theory seem not to be relevant for FKM if no dispersion of localized electron is introduced.

\section{\label{sec5:level1}Conclusion}

In conclusion, our study provides observations of QOs within the strongly correlated, doped FKM. In the PS state, we discern clear SdH oscillations and notable oscillations in electron density $n_c$. In contrast, observables in the low-temperature regime of NFL phases reveals only a weak, irregular magnetic field dependence.
As the hole concentration increases, the Fermi surface undergoes a transition from a hole-like to an electron-like type, with the QOs frequency in both instances adhering to the Onsager relation. The temperature-dependent amplitude behavior aligns well with the extended LK formula when an appropriate renormalization of the effective mass is employed.
In comparison with the Fermi liquid system, the substantial imaginary part of the self-energy in the strongly-correlated FKM leads to a global enhancement of the effective mass. This enhancement accounts for the observed suppression of high-temperature QOs. The low-temperature CDW phase, which witnesses a gap formation around the Fermi energy, exhibits no regular QO signal.
Conversely, under weak correlations where the interaction-induced gap is minimal, the insulating CDW state reveals distinct QOs with frequencies corresponding to the metallic Fermi surface prior to gap formation. Analytical calculations illuminate the origin of these QOs within the insulating state, demonstrating that the gap effect is akin to the Dingle factor, thus leading to suppression of QOs as the correlation strength increases.
These observations, coupled with our previous work on doped FKM that uncovered a strong interaction-induced violation of the Luttinger theorem, underscore the pivotal role of quasiparticles in the manifestation of regular QOs, as evidenced in the PS state. Moreover, the QOs in strongly correlated systems align well with the LK theory, providing an analytical depiction of the self-energy exists. In this point, we have noticed recent studies on QOs in Hatsugai-Kohmoto model, where non-Landau quasiparticle still leads to QOs but non-LK behaviors has been extracted \cite{PhysRevB.108.085106,zhong2023notes}.
Collectively, our findings illuminate the nature of QOs in strongly correlated systems and underscore the significance of quasiparticles in their emergence and description.

	\begin{acknowledgements}
This research was supported in part by Supercomputing Center of
Lanzhou University and NSFC under Grant No.~$11834005$, No.~$11874188$.	
We thank the Supercomputing Center of Lanzhou University for allocation of CPU time.		
	\end{acknowledgements}

\bibliography{ref}



\end{document}